\documentclass[runningheads]{llncs}
\usepackage[T1]{fontenc}
\usepackage{graphicx}
\usepackage{subcaption}
\usepackage{multirow}
\usepackage{tabularx}
\usepackage{makecell}
\usepackage{booktabs}
\usepackage{url}

\begin{document}
\title{Investigating Multimodal Large Language Models to Support Usability Evaluation}
\titlerunning{Investigating Multimodal LLMs to Support Usability Evaluation}
%
\author{Sebastian Lubos\inst{1}\orcidID{0000-0002-5024-3786} \and
Alexander Felfernig\inst{1}\orcidID{0000-0003-0108-3146} \and
Damian Garber\inst{1}\orcidID{0009-0005-0993-0911} \and
Gerhard Leitner\inst{2}\orcidID{0000-0002-3084-0727} \and
Julian Schwazer\inst{2}\orcidID{0009-0004-6431-5033} \and
Manuel Henrich\inst{3}\orcidID{0009-0007-4644-0836}
}

\authorrunning{S. Lubos et al.}

\institute{Graz University of Technology, Inffeldgasse 16b/II, 8010 Graz, Austria
\email{sebastian.lubos@tugraz.at} \and
University of Klagenfurt, Universitätsstraße 65-67, 9020 Klagenfurt, Austria \and
UNiQUARE Software Development GmbH, Lannerweg 9, 9201 Krumpendorf, Austria
}

%
%
\maketitle              
\begin{abstract}
\textit{Usability evaluation} is an essential method to support the design of effective and intuitive \textit{user interfaces (UIs)}. However, it commonly relies on resource-intensive, expert-driven methods, which limit its accessibility, especially for small organizations. 
Recent \textit{multimodal large language models (MLLMs)} have the potential to support usability evaluation by analyzing textual instructions together with visual UI context. This paper investigates the use of MLLMs as assistive tools for usability evaluation by framing the task as a prioritization problem. It identifies and explains usability issues and ranks them by severity. We report a study that compares the evaluations generated by multiple MLLMs with assessments from usability experts. The results demonstrate that MLLMs can offer complementary insights and support the efficient prioritization of critical issues. Additionally, we present an interactive visualization tool that enables the transparent review and validation of model-generated findings. Based on this, we outline concepts for integrating MLLM-based usability evaluation into real-world development workflows. 

\end{abstract}

\section{Introduction}\label{sec:introduction}
The \textit{usability} of an application, website, or other software system interface can be described by a set of quality attributes that determine how effectively users interact with it~\cite{winter2008comprehensive}. Good usability is widely recognized as a critical success factor, as it directly influences whether users can find, understand, and use system functionality~\cite{iso2018iso9241}. Ensuring usability requires systematic evaluation to identify user difficulties and improve usability and \textit{user experience (UX)}~\cite{hass2019practical,holingsed2007usability}. 

However, conventional evaluation methods such as usability \textit{testing} and expert \textit{inspections} are time-consuming, costly, and rely on specialized expertise. To reduce this effort, recent research has investigated the use of \textit{multimodal large language models (MLLMs)} to support usability evaluation~\cite{duan2024generating,guerino2025can,lubos2025towards,pourasad2025does}. Unlike text-only models, MLLMs can jointly process textual and visual input~\cite{geminiteam2024gemini}. For usability analysis, this allows UI screenshots to be provided together with explicit usability evaluation instructions. Initial studies report promising overlap with expert assessments~\cite{pourasad2025does}, which suggests that MLLMs can support semi-automated usability evaluation with human validation.

A fundamental characteristic of usability evaluation is the absence of a unique ground truth, as even expert evaluators often disagree on which issues are present and how severe they are~\cite{molich2018are}. While this variability complicates evaluation, it also suggests that different evaluators (human or model-based) may reveal complementary usability issues of a UI. To investigate this aspect, we compare multiple MLLMs to analyze the diversity and overlap of identified usability issues and their alignment with assessments of usability experts.

In practice, usability evaluations often result in extensive lists of issues that must be reviewed and prioritized. This creates an additional challenge, as developers must decide which problems deserve their limited attention and resources. We therefore frame automated usability evaluation as a \textit{prioritization problem} and investigate whether MLLMs can not only identify but also rank usability issues by severity. Such support could make usability evaluation more efficient and accessible, particularly for small teams that lack dedicated UX expertise.

To this end, we conduct a comparative study of MLLMs for automated usability evaluation and introduce an interactive visualization tool that aggregates and presents detected issues for review. Based on our findings, we discuss design implications and outline directions for scalable and cost-effective usability evaluation using MLLMs.
The main contributions of this paper are the following:
\begin{enumerate}
    \item We formalize an approach for semi-automated usability evaluation using MLLMs that prioritizes usability issues by severity.
    \item We report results of a \textit{proof-of-concept study} that compares MLLM-generated evaluations with assessments of usability experts.
    \item We present a visualization tool for reviewing usability issues and outline considerations for integrating MLLM-based analysis in development workflows.
\end{enumerate}

\section{Background and Related Work}\label{sec:background}
Usability is a core concept in \textit{human–computer interaction (HCI)} and is commonly defined as the ``\textit{extent to which a product can be used by specified users to achieve specified goals with effectiveness, efficiency and satisfaction in a specified context of use.}''~\cite{iso2018iso9241}. Accordingly, \textit{usability evaluation} is applied throughout the entire \textit{software development lifecycle}~\cite{ruparelia2010software} to identify issues and suggest improvements. Established evaluation approaches include \textit{usability testing} with end users~\cite{hass2019practical} and \textit{usability inspections} conducted by experts who imitate real users~\cite{holingsed2007usability}. While both approaches are effective, they are resource-intensive and depend on the active participation of end users or usability experts.

Structured systematic evaluation uses established methods such as the \textit{Nielsen heuristics}~\cite{nielsen1994enhancing} and \textit{cognitive walkthroughs}~\cite{spencer2000streamlined}. These methods help identify common usability issues and task-related barriers, but still require substantial manual effort and domain expertise. As a result, they are not consistently applied in practice, particularly in small teams or early development phases.

To reduce the evaluation effort, prior work has explored automated and AI-based usability evaluation techniques. Earlier approaches relied on rule-based tools and heuristic checkers~\cite{castro2022automated,namoun2021review}, which typically address only narrow usability aspects and struggle with subjective or context-dependent issues~\cite{kuric2025systematic}. Recently, \textit{multimodal large language models (MLLMs)} offer new opportunities by combining textual and visual input for a more comprehensive understanding of UIs~\cite{geminiteam2024gemini}. 

Recent work uses MLLMs to analyze application screenshots alongside explicit evaluation instructions and usability criteria~\cite{duan2024generating,guerino2025can,lubos2025towards,pourasad2025does}. These studies indicate their potential to complement human reviews and report substantial overlap between MLLM-generated analyses and expert usability assessments, though human validation remains necessary~\cite{guerino2025can,pourasad2025does}. 

In contrast to prior work, we frame usability evaluation as a \textit{prioritization problem}. Our approach aggregates and ranks identified issues by severity to help decide which problems should be addressed first. We further introduce an interactive visualization tool that enables transparent review and human validation. Together, these elements position MLLMs as scalable assistants for usability evaluation, particularly for developers and teams with limited UX expertise.

\section{Multimodal LLMs for Usability Evaluation}\label{sec:mllm-based}
Building on the concept of human–AI collaboration in usability evaluation, our approach uses MLLMs to help developers identify and prioritize usability issues. The MLLM processes textual and visual inputs, including interface screenshots, persona descriptions (i.e., brief descriptions of representative user roles and goals), and evaluation criteria, to detect contextually relevant issues and assign severity scores. These identified issues are ranked and accompanied by explanations of the underlying reasoning to support informed decision-making, particularly for developers without formal usability expertise.

\begin{figure}[t]
    \centering
    \includegraphics[width=0.80\linewidth]{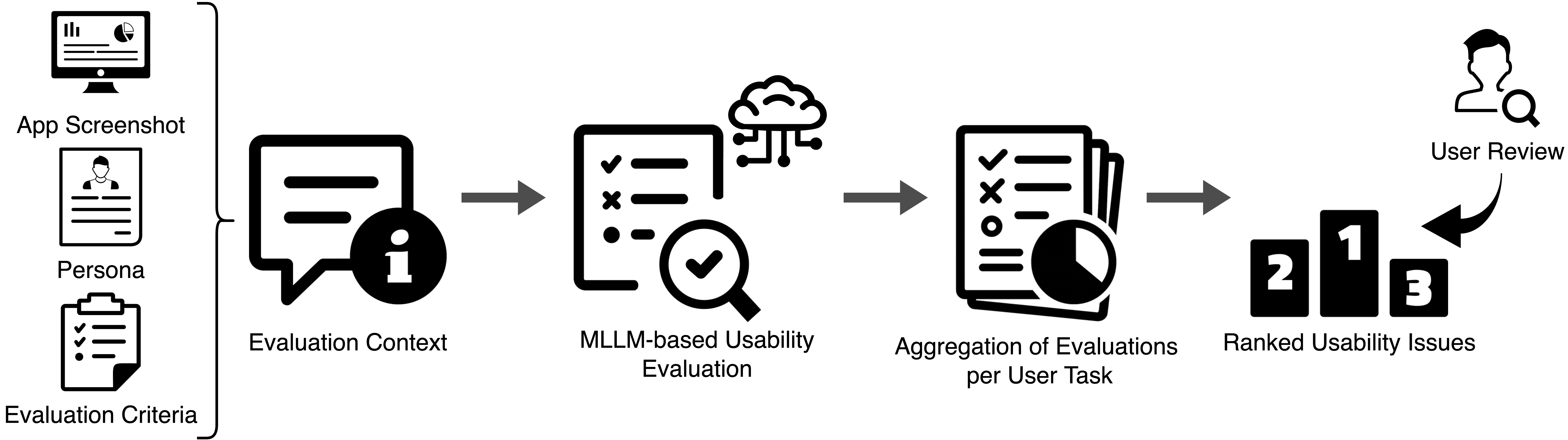}
    \vspace{-4pt}
    \caption{Overview of the MLLM-based usability evaluation workflow, in which usability issues are identified and ranked by severity for user review.}
    \label{fig:overview}
    \vspace{-8pt}
\end{figure}

\begin{figure}
    \centering
    \includegraphics[width=0.75\linewidth]{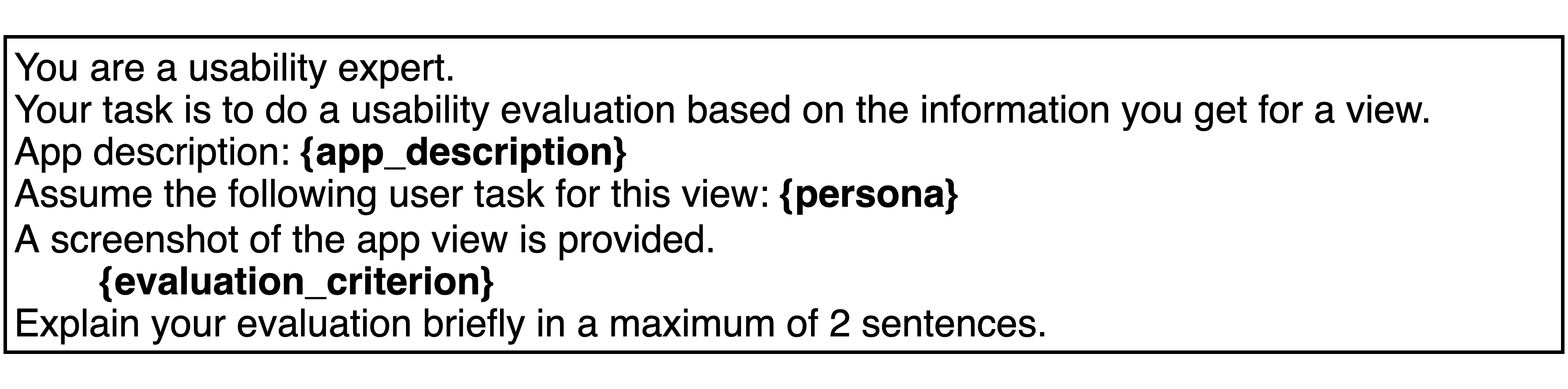}
    \vspace{-8pt}
    \caption{Usability evaluation prompt template with variables shown in braces ``\textbf{\{\}}''.}
    \label{fig:prompt}
    \vspace{-16pt}
\end{figure}

Figure~\ref{fig:overview} illustrates the evaluation workflow. The \textit{evaluation context} comprises an application screenshot, a persona description, and a selected evaluation criterion. Although multiple criteria could be evaluated within a single prompt, we restrict each prompt to a single criterion to improve the precision of assessments and reduce hallucinations. The inputs are used in the prompt template (see Figure~\ref{fig:prompt}) to guide the MLLM in the UI analysis from a specified user perspective.

When a user task spans multiple UI states, each screenshot is evaluated independently for the same criterion. The severity of a given criterion is aggregated at the task level using the highest observed severity. This conservative strategy ensures that critical issues are not hidden by less severe interface states.

To structure the analysis of an application, we organize evaluations around user tasks that represent concrete usage scenarios. At the task level, the overall severity is determined by the most severe issue identified across all associated criteria and screenshots. Tasks with the same maximum severity are further distinguished by the number of issues at that level. The resulting task prioritization highlights the user tasks that are most likely to require attention. Developers can review the identified issues and MLLM explanations to validate or refine the results based on their domain knowledge. This semi-automated process reduces the effort required for exhaustive manual evaluations while preserving human control over interpretation and prioritization.

\section{Visualization of Usability Issues}\label{sec:visualization}
\begin{figure}[t]
    \centering
    \includegraphics[width=0.75\linewidth]{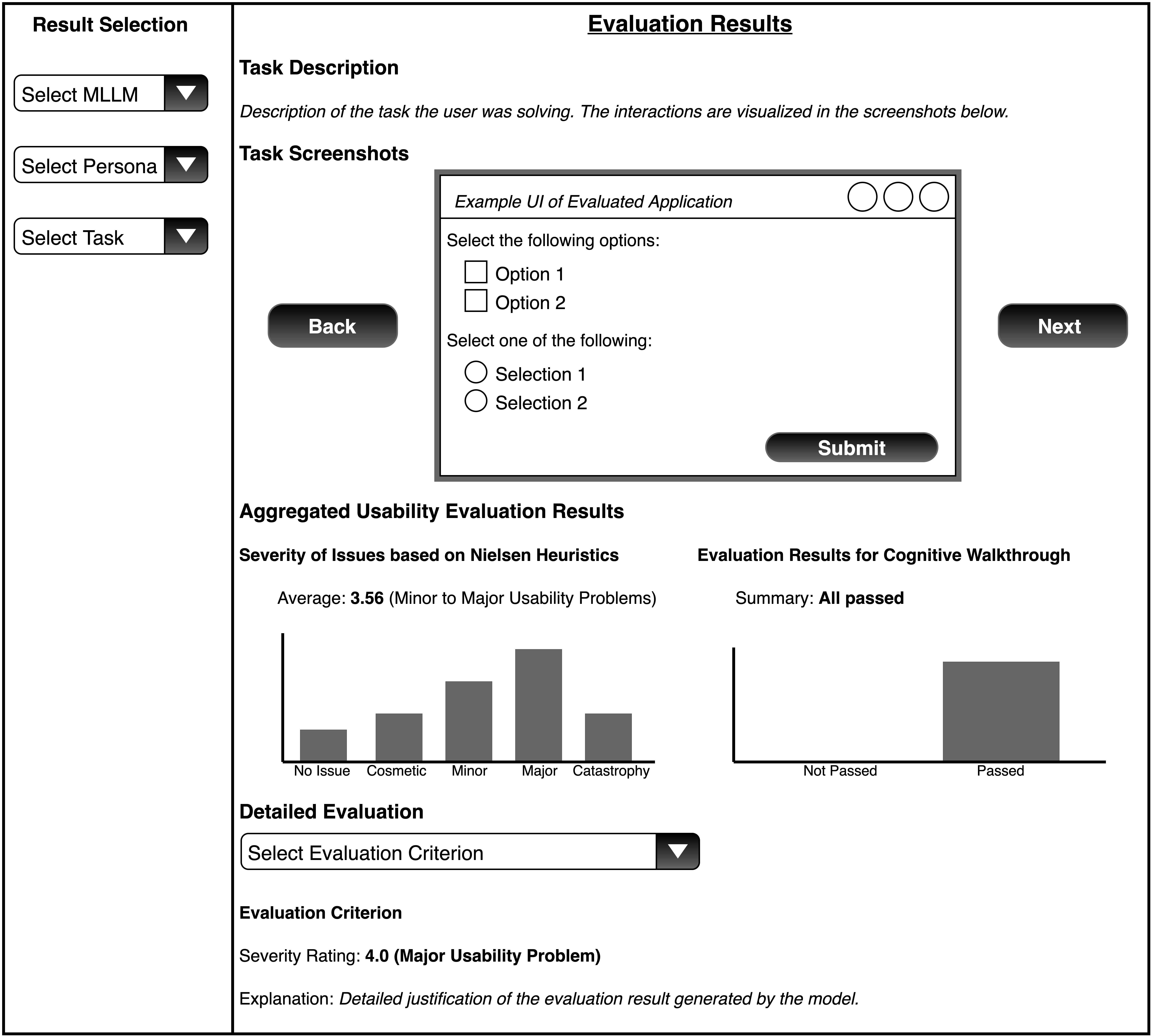}
    \vspace{-6pt}
    \caption{Wireframe of the visualization tool. Users select an LLM, persona, and task (left) to review the aggregated and detailed evaluation results (right).}
    \label{fig:Mockup}
    \vspace{-16pt}
\end{figure}

To support the review and interpretation of MLLM-generated usability evaluations, we developed an interactive visualization tool using \textit{Streamlit}\footnote{\url{https://streamlit.io}}. The tool enables users to explore evaluation results across different MLLMs, personas, and user tasks and presents identified usability issues in a transparent and concise manner. It supports both task-level overviews and detailed inspections of individual criteria, as well as MLLM explanations. Figure~\ref{fig:Mockup} shows the tool using a wireframe. The fully implemented application is available in our repository.\footnote{\url{https://github.com/AIG-ist-tugraz/Compare-MLLM-Usability-Evaluation}}

The visualization supports three hierarchical levels of exploration: (\textit{i}) MLLM selection, (\textit{ii}) task-level prioritization, and (\textit{iii}) criterion-level inspection. Users first select an MLLM, persona, and task to load the corresponding evaluation results. Each task view includes a brief description and one or more screenshots that represent the interaction sequence. For tasks spanning multiple UI states, screenshots can be browsed interactively to support understanding of the user flow and to relate identified issues to specific UI elements.

At the task level, the visualization summarizes the aggregated severity derived from the most critical issues identified across associated screenshots and evaluation criteria. Separate bar charts show the distribution of \textit{Nielsen heuristic} severity scores and \textit{cognitive walkthrough} outcomes for the selected task. For more detailed inspection, users can select individual evaluation criteria to see assigned severity ratings and the MLLM-generated explanation.

\section{Methodology}\label{sec:methodology}
Our study investigates whether MLLMs can support usability evaluation by identifying and prioritizing usability issues. We compare MLLM-generated assessments with expert evaluations as ground truth using a task-centered pipeline. Although the severity is assessed per evaluation criterion, all quantitative analyses are conducted at the \textit{task level}. The goal is to prioritize user tasks that involve the most critical usability issues. The evaluation comprises four steps:

\begin{enumerate}
    \item Usability experts evaluate predefined tasks and document usability issues. They capture screenshots of involved UI states (views).
    \item MLLMs generate usability evaluations for the same tasks and screenshots.
    \item The view- and criterion-level severity scores are aggregated to obtain task-level usability issue severity and rankings.
    \item MLLM outputs are compared with expert assessments using agreement and ranking metrics.
\end{enumerate}

\subsection{Expert Evaluations}
Two experienced usability experts\footnote{One expert has over 20 years of experience in usability engineering and HCI/UX, while the other specializes in multimodal user interfaces and intelligent interactive systems for human–machine collaboration.} independently evaluated \textit{KnowledgeCheckR},\footnote{\url{https://www.knowledgecheckr.com/?lang=en}} a web-based learning platform used in educational and corporate contexts. The system supports, for example, quiz creation, participation, and result review.

The experts followed predefined usage scenarios but were free to choose their interaction paths, which reflects common usability inspection practice. Both experts evaluated two personas: (\textit{i}) a \textit{teacher} creating a quiz and (\textit{ii}) a \textit{student} completing a quiz. Because the interaction paths were not identical, we cannot directly compare the evaluations of the two experts. Therefore, we treated these evaluations as independent expert perspectives.

Both experts applied \textit{Nielsen heuristics} and \textit{cognitive walkthroughs} using consistent rating schemes (ordinal school grades for heuristics and binary decisions for walkthroughs). During the evaluations, they captured screenshots to document relevant UI states. These screenshots were reused as inputs for the MLLM-based analysis to ensure a consistent evaluation context.

A \textit{persona} represents a user role and overarching goal, which is decomposed into concrete \textit{tasks} (e.g., creating a quiz). Each task involves one or more \textit{views}, corresponding to individual UI states captured as screenshots. Depending on task complexity, the tasks comprised between one and seven views. Across both personas, the evaluation covered $15$ tasks and $52$ views.

\subsection{MLLM-Based Analysis}
We provided the screenshots collected during the expert evaluations to the MLLMs together with persona descriptions and evaluation criteria (see Section~\ref{sec:mllm-based}). This resulted in two parallel MLLM-based analyses of the same system, each aligned with one expert’s interaction path. We used six MLLMs (see Table~\ref{tab:llm_overview}) for the analyses, including \textit{general-purpose} and \textit{reasoning-oriented} models of different sizes. As the exact model capacities are undisclosed, we categorize them by type and relative scale. If supported by the models, \textit{temperature} was set to $0$ to promote more deterministic output. 

\begin{table}[h!]
    \vspace{-24pt}
    \centering
        \caption{Overview of MLLMs used in our experiments.}
    \label{tab:llm_overview}
    \begin{tabular}{@{}llllc@{}}
        \toprule
        \textbf{LLM} & \textbf{Version} & \textbf{Developer} & \textbf{Category} & \textbf{Reference} \\
        \midrule
        gemini-2.0-flash-lite & 001 & Google & small, general & \cite{geminiteam2024gemini} \\
        gemini-2.5-pro & preview-03-25 & Google & large, general & \cite{geminiteam2024gemini} \\
        gemini-2.0-flash-thinking & exp-01-21 & Google & reasoning & \cite{geminiteam2024gemini} \\
        gpt-4.1-nano & 2025-04-14 & OpenAI & small, general & \cite{openai2025gpt41} \\
        gpt-4.1 & 2025-04-14 & OpenAI & large, general & \cite{openai2025gpt41} \\
        o1 & 2024-12-17 & OpenAI & reasoning & \cite{openai2024openaio1card} \\
        \bottomrule
    \end{tabular}
    \vspace{-24pt}
\end{table}

\subsection{Usability Issue Aggregation and Prioritization}
To support prioritization at the task level, we apply a conservative aggregation strategy. For each task, the overall issue severity is defined as the \textit{maximum severity} observed across all associated screenshots and evaluation criteria. For \textit{Nielsen heuristics}, this corresponds to the highest ordinal severity rating. For \textit{cognitive walkthroughs}, a task is considered failed if any walkthrough step fails. This reflects the practical assumption that a single severe issue is sufficient to warrant attention. When multiple tasks share the same maximum severity, ties are resolved by the number of issues at that severity level for a stable ordering.

\vspace{-4pt}
\subsection{Evaluation Metrics}
We evaluate the MLLM-based evaluations quantitatively to address two goals:
\begin{itemize}
    \item[] \textbf{Rating agreement}: The task-level severity ratings are compared between MLLMs and experts. For ordinal \textit{Nielsen heuristic} ratings, we compute the \textit{mean absolute deviation (MAD)}, which quantifies how far the MLLM ratings deviate from expert judgments. For binary \textit{cognitive walkthrough} scores, we measure \textit{accuracy}, defined as the proportion of tasks for which the MLLM and the expert assign the same passed/failed outcome.
    \vspace{2pt}
    \item[] \textbf{Prioritization quality}: To measure whether the most severe usability issues identified by experts appear among the top-ranked MLLM results, we compare rankings derived from aggregated severity scores using \textit{Precision@k}.
\end{itemize}

Finally, we conduct a qualitative comparison of expert and MLLM-generated explanations to assess whether they reference similar UI elements and usability principles. This provides insight into the alignment of human and model reasoning beyond numerical agreement.

\section{Results}\label{sec:evaluationresults}
The evaluation included $15$ explicit user tasks embedded in persona-specific usage scenarios. Each task aggregates usability issues encountered across one or more UI views during task execution. All results are reported at the \textit{task level}, as our primary goal is to assess whether MLLMs can prioritize which user tasks are affected by critical usability issues in alignment with expert judgments.

\vspace{2pt}
\textbf{Task-level severity agreement}:
For the \textit{Nielsen heuristics}, the \textit{MAD} values (see Table~\ref{tab:mad}) show that MLLM-generated severity ratings deviate by less than one severity level from expert judgments on average. However, the numbers vary across MLLMs and experts. Overall, the larger (\textit{gemini-2.5-pro} and \textit{gpt-4.1}) and reasoning-oriented MLLMs (\textit{gemini-2.0-flash-thinking} and \textit{o1}) had lower deviations than the smaller models. As severity assessments are subjective, the agreement metrics mainly indicate whether the MLLM severity scores fall within the range of expert judgments and reflect relative task criticality.

\begin{table}[h!]
    \vspace{-11pt}
    \centering
    \caption{\textit{Mean Absolute Deviation (MAD)} between expert and MLLM severity ratings for \textit{Nielsen heuristics}. Lower values indicate closer alignment.}
    \label{tab:mad}
    \vspace{2pt}
    \begin{tabular}{@{}lcc|c@{}}
        \toprule
        \textbf{MLLM} & $exp_1$ & $exp_2$ & \textit{both experts} \\
        \midrule
        \textit{gemini-2.0-flash-lite} & 0.78 & 1.50 & 1.07 \\
        \textit{gemini-2.5-pro} & \textbf{0.22} & 0.83 & 0.47\\
        \textit{gemini-2.0-flash-thinking} & \textbf{0.22} & \textbf{0.67} & \textbf{0.40} \\
        \textit{gpt-4.1-nano} & 1.33 & 2.0 & 1.60 \\
        \textit{gpt-4.1} & 0.67 & 0.83 & 0.73\\
        \textit{o1} & \textbf{0.22} & 1.50 & 0.73\\
        \midrule
        \textit{All MLLMs} & 0.57 & 1.22 & 0.83 \\
        \bottomrule
    \end{tabular}
    \vspace{-16pt}
\end{table}

\begin{table}[h!]
    \vspace{-0pt}
    \centering
    \caption{\textit{Accuracy} between expert and MLLM task-level \textit{passed/failed} judgments for \textit{cognitive walkthroughs}. Higher values indicate closer alignment.}
    \label{tab:accuracy}
    \vspace{2pt}
    \begin{tabular}{@{}lcc|c@{}}
        \toprule
        \textbf{MLLM} & $exp_1$ & $exp_2$ & \textit{both experts} \\
        \midrule
        \textit{gemini-2.0-flash-lite} & 0.67 & \textbf{0.83} & 0.73 \\
        \textit{gemini-2.5-pro} & \textbf{0.78} & 0.50 & 0.67 \\
        \textit{gemini-2.0-flash-thinking} & \textbf{0.78} & \textbf{0.83} & \textbf{0.80} \\
        \textit{gpt-4.1-nano} & 0.22 & 0.50 & 0.33 \\
        \textit{gpt-4.1} & 0.0 & 0.67 & 0.27 \\
        \textit{o1} & 0.11 & 0.50& 0.27 \\
        \midrule
        \textit{All MLLMs} & 0.43 & 0.64 & 0.51 \\
        \bottomrule
    \end{tabular}
    \vspace{-18pt}
\end{table}

For the \textit{cognitive walkthroughs}, Table~\ref{tab:accuracy} reports the \textit{accuracy} between MLLMs and experts. The \textit{gemini} MLLMs had higher accuracy than the \textit{OpenAI} models. \textit{gemini-2.0-flash-thinking} achieved the highest overall accuracy in matching the experts \textit{passed/failed} judgments. This alignment suggests that several MLLMs can reliably identify tasks where usability issues prevent successful completion.

\vspace{2pt}
\textbf{Prioritization of critical tasks}:
To evaluate how well MLLMs prioritize critical usability issues, we compare expert- and MLLM-based rankings of tasks involving usability issues using \textit{Precision@k} (see Table~\ref{tab:precision}). Critical usability issues are defined as those that substantially affect task completion. While none of the MLLMs consistently identified the single most critical task ($k=1$), several achieved good precision for larger values of $k$. This indicates meaningful alignment with expert judgments when prioritizing sets of tasks affected by critical usability issues, even when exact top-ranked agreement is unstable.

For \textit{Nielsen heuristics}, the reasoning-oriented model \textit{o1} delivered the most consistent prioritization. For \textit{cognitive walkthroughs}, multiple MLLMs prioritized critical tasks similarly to experts. This suggests that the explicit, task-centered evaluation criteria reduce ambiguity. Overall, these results reinforce that task-level prioritization, rather than identifying a single most severe task or predicting exact severity scores, is more robust and practically relevant.

\begin{table}[h!]
    \vspace{-11pt}
    \centering
    \caption{\textit{Precision@k} of task-level prioritization between experts and MLLMs for \textit{Nielsen heuristics} and \textit{cognitive walkthroughs}. Higher values indicate better agreement in prioritizing tasks affected by critical usability issues.}
    \label{tab:precision}
    \vspace{2pt}
    \begin{tabular}{@{}lccc|ccc||ccc|ccc@{}}
        \toprule
        \multirow{3}{6em}{\textbf{MLLM}} & \multicolumn{6}{c||}{\textit{Nielsen heuristics}} & \multicolumn{6}{c}{\textit{Cognitive walkthrough}} \\
        & \multicolumn{3}{c|}{$exp_1$} & \multicolumn{3}{c||}{$exp_2$} & \multicolumn{3}{c|}{$exp_1$} & \multicolumn{3}{c}{$exp_2$}\\
        & $@2$ & $@3$ & $@5$ & $@2$ & $@3$ & $@5$ & $@2$ & $@3$ & $@5$ & $@2$ & $@3$ & $@5$ \\
        \midrule
        \textit{gemini-2.0-flash-lite} & 0.0 & 0.0  & 0.40 & 0.0 & 0.33 & \textbf{0.80} & \textbf{0.50} & 0.67  & \textbf{0.80} & 0.50 & \textbf{0.67} & \textbf{1.0}\\
        \textit{gemini-2.5-pro} & 0.0 & 0.33  & \textbf{0.60} & 0.0 & 0.0 & \textbf{0.80} & \textbf{0.50} & 0.67 & 0.60 & 0.0 & 0.33 & 0.80 \\
        \textit{gemini-2.0-flash-think.} & 0.0 & 0.33  & \textbf{0.60} & 0.0 & 0.33 & \textbf{0.80} & \textbf{0.50} & \textbf{1.0}  & \textbf{0.80} & 0.50 & \textbf{0.67} & \textbf{1.0} \\
        \textit{gpt-4.1-nano} & 0.0 & 0.0  & 0.40 & 0.0 & 0.33 & \textbf{0.80} & \textbf{0.50} & 0.33 & \textbf{0.80} & 0.50 & 0.33 & \textbf{1.0} \\
        \textit{gpt-4.1} & 0.0 & 0.0  & 0.40 & 0.0 & 0.33 & \textbf{0.80} & 0.0 & 0.33 & \textbf{0.80} & \textbf{1.0} & \textbf{0.67} & \textbf{1.0} \\
        \textit{o1} & \textbf{0.5} & \textbf{0.67} & \textbf{0.60} & \textbf{0.50} & \textbf{0.67} & \textbf{0.80} & 0.0 & 0.33 & \textbf{0.80} & 0.50 & 0.33 & 0.80 \\
        \bottomrule
    \end{tabular}
    \vspace{-14pt}
\end{table}

\vspace{2pt}
\textbf{Qualitative insights}:
To complement the quantitative metrics and better understand how MLLMs reason about usability issues, we compare expert and MLLM-generated explanations (see examples in Table~\ref{tab:examples}). This analysis shows that the MLLMs identify and describe similar visually grounded usability issues encountered during task execution, for example, the visibility and labeling of UI elements. However, MLLMs frequently missed issues that depend on the interaction dynamics or temporal context, which are not captured in static screenshots. The examples further reveal that different MLLMs may draw different conclusions from the same visual context. This suggests that combining outputs across models, for example, through majority voting or consensus-based strategies, and providing richer contextual input (e.g., screen recordings of interactions) could improve the reliability and coverage of MLLM-based usability evaluations.

\begin{table}[t]
    \vspace{-12pt}
    \centering
    \caption{Comparison of expert and MLLM explanations for usability issues.}
    \label{tab:examples}
    \vspace{2pt}
    \begin{tabular}{@{}p{9.7em}p{15.0em}p{12.2em}@{}}
        \toprule
        \textbf{Expert} & \textbf{LLM} & \textbf{Observations}\\
        \midrule
        \textit{The goal of sharing the created quiz with a student is reasonable [...], especially since the key and share buttons are prominently displayed.} & \textbf{gemini-2.0-flash-lite}: \textit{The view provides clear options to add new questions, and the existing questions are displayed with their types [...]. The teacher can easily add new questions and share the quiz.}  & Both explanations mention quiz sharing, though the LLM places less emphasis on it. More detailed context or cropped screenshots could improve relevance.  \\
        \midrule
        \textit{The user might have difficulties identifying the correct selection on the dialogue. In the current task, this would be ``Knowledge-Check'' [...]} & \textbf{gemini-2.0-flash-lite}: \textit{The user might not easily identify the ``Knowledge-Check'' option as the correct action to start the quiz, leading to confusion.}\newline\textbf{gemini-2.0-flash-thinking}: \textit{The ``Knowledge-Check'' button is visible and labeled appropriately for accessing the quiz, so the student will likely notice the correct action.} & All explanations mention the ``Knowledge-Check'' option. However, one LLM and the expert indicate potential user confusion, while the other assumes clarity. This variation suggests that a majority vote could enhance assessment reliability. \\
        \midrule
        \textit{Once the user enters an input field, the entire description help [...] is gone, and they must remember which information they are currently inputting.} & \textbf{o1}: \textit{Important fields (e.g., quiz name, question description, question type icons) are visible, reducing users' memory load. However, more in-context help (e.g., tooltips for question-type icons) would further clarify available options and improve recognition.}  & This shows a current limitation of the LLM-based evaluation. The expert identifies an interaction issue that is not visible in a static screenshot. More dynamic input (e.g., multiple frames) could help capture such issues. \\
        \bottomrule
    \end{tabular}
    \vspace{-18pt}
\end{table}

\vspace{2pt}
\textbf{Summary}:
Overall, the results indicate that MLLMs can support usability evaluation by identifying the most critical usability issues. While the MLLM-generated severity scores fall within the range of expert judgments, the exact reproduction of expert severity ratings remains less reliable. However, they support the ranking of usability issues by severity and align well with expert assessments. The qualitative analysis further reveals that MLLMs effectively capture many visually grounded usability problems but remain limited in addressing interaction-dependent issues. This emphasizes the role of MLLMs as supportive tools for usability analysis in practice, rather than replacements for experts.

\section{Practical Implications for Usability Analysis}\label{sec:integration}
Our results suggest that MLLM-based usability analysis can provide practical value when used to support the prioritization of usability issues. This has implications for how such approaches can be integrated into development workflows.

\vspace{2pt}
\textbf{Usability issue review and collaboration}:
The observed alignment between MLLMs and experts in prioritizing critical usability issues indicates that MLLMs can support early-stage usability analysis. Ranked issue lists help developers and designers to quickly identify which user tasks warrant closer inspection, for example, during sprint planning or design reviews. The effort for a manual review of the entire application is reduced. Our visualization tool (see Section~\ref{sec:visualization}) supports this process by linking prioritized issues to screenshots and concise explanations across different MLLMs. Extending such visualizations, for example, by highlighting agreement and disagreement across models or by aggregating results, could further support collaborative review, validation of findings, and prioritization of usability improvements for release planning.

\vspace{0pt}
\textbf{Evaluation context}:
The qualitative analysis shows that MLLMs reliably capture visually grounded usability issues but often miss interaction-dependent problems that depend on dynamic UI changes. This suggests that providing richer evaluation context, such as short interaction recordings, usage logs, or source code, could improve the identification of such issues. With access to implementation-level context, MLLMs could further support usability analysis by suggesting explicit code-level improvements.

\vspace{2pt}
\textbf{Scalability, and privacy}:
Manual usability evaluations require active participation from experts, which makes it difficult to conduct them continuously throughout development. MLLM-based usability analysis can enable more frequent evaluations, which allows ongoing reflection on usability across different development phases. However, practical deployments require careful handling of sensitive data, as screenshots and design artifacts may contain confidential information. Supporting on-premise execution and secure data handling is therefore essential to balance scalability and privacy requirements in the industry.

\vspace{-3pt}
\section{Limitations}\label{sec:limitations}
\vspace{-2pt}
Firstly, the evaluation was conducted with two usability experts and a single application, which limits the generalizability of the results. While the system represents a realistic use case, future studies should consider additional applications, domains, and evaluators.
Secondly, MLLM performance depends on model choice and prompt design. Although we applied a consistent prompt structure across models, variations in prompts or parameters may influence the identified issues and their prioritization.
Finally, usability evaluation is inherently subjective, and expert assessments do not constitute an objective ground truth. While we mitigated this by using two independent expert references, future work could incorporate larger evaluator groups or consensus-based reference data to further enhance the reliability of the results.

\section{Conclusion and Future Work}\label{sec:conclusion}
In this paper, we investigated the use of \textit{multimodal large language models (MLLMs)} to support usability evaluation by identifying, prioritizing, and explaining usability issues based on user interface screenshots and contextual information. The results show that MLLMs can reliably prioritize critical usability issues in many cases. This suggests that MLLMs are well-suited as assistive tools for semi-automated usability analysis to support the early identification of usability issues, particularly for teams with limited usability expertise. Our presented visualization tool further demonstrates how MLLM outputs can be integrated into human-centered review processes to enable transparent inspection.

For future work, we plan to explore richer interaction contexts beyond static screenshots to better capture interaction-dependent usability issues. Additionally, we plan to investigate the integration of MLLM-based usability analysis into real-world development workflows to assess its practical impact.

\begin{credits}
\subsubsection{\ackname} The presented work has been developed within the research project \textsc{GENRE}, which is funded by the Austrian Research Promotion Agency (FFG) under the project number \textsc{915086}.

\subsubsection{\discintname}
The authors have no competing interests to declare that are
relevant to the content of this article.
\end{credits}

%
%
%
\bibliographystyle{splncs04}
\bibliography{bibliography}

\end{document}